\documentclass[aps,preprint,superscriptaddress]{revtex4-1}

\usepackage{hyperref}
\hypersetup{
    bookmarks=true,         % show bookmarks bar?
    unicode=true,          % non-Latin characters in Acrobatâs bookmarks
    pdftoolbar=false,        % show Acrobatâs toolbar?
    pdfmenubar=true,        % show Acrobatâs menu?
    pdffitwindow=true,      % page fit to window when opened
    pdftitle={My title},    % title
    pdfauthor={Author},     % author
    pdfsubject={Subject},   % subject of the document
    pdfnewwindow=true,      % links in new window
   pdfkeywords={keywords}, % list of keywords
   colorlinks=true,       % false: boxed links; true: colored links
    linkcolor=black,          % color of internal links
    citecolor=black,        % color of links to bibliography
    filecolor=magenta,      % color of file links
    urlcolor=cyan           % color of external links
}

\usepackage{graphicx}
\usepackage{amsmath}
\usepackage{natbib}
\usepackage{layouts}

\begin{document}

%\title{Configurational disorder effects on adatom surface mobilities of Ti$_{1-x}$Al$_{x}$N alloys}
\title{Effects of configurational disorder on adatom mobilities on Ti$_{1-x}$Al$_{x}$N(001) surfaces}

\author{B. Alling} 
\email{bjoal@ifm.liu.se}
\affiliation{Department of Physics, Chemistry and Biology (IFM), Link\"oping University, SE-581 83  Link\"oping, Sweden}
\author{P. Steneteg}
\affiliation{Department of Physics, Chemistry and Biology (IFM), Link\"oping University, SE-581 83 Link\"oping, Sweden}
\author{C. Tholander}
\affiliation{Department of Physics, Chemistry and Biology (IFM), Link\"oping University, SE-581 83 Link\"oping, Sweden}
\author{F. Tasn\'adi}
\affiliation{Department of Physics, Chemistry and Biology (IFM), Link\"oping University, SE-581 83 Link\"oping, Sweden}
\author{I. Petrov}
\affiliation{Department of Physics, Chemistry and Biology (IFM), Link\"oping University, SE-581 83 Link\"oping, Sweden}
\affiliation{Frederick Seitz Materials Research Laboratory and the Materials Science Department, University of Illinois at Urbana-Champaign, Urbana, Illinois 61801, USA}
\author{J. E. Greene}
\affiliation{Department of Physics, Chemistry and Biology (IFM), Link\"oping University, SE-581 83 Link\"oping, Sweden}
\affiliation{Frederick Seitz Materials Research Laboratory and the Materials Science Department, University of Illinois at Urbana-Champaign, Urbana, Illinois 61801, USA}
\author{L. Hultman}
\affiliation{Department of Physics, Chemistry and Biology (IFM), Link\"oping University, SE-581 83 Link\"oping, Sweden}

\date{\today}

\begin{abstract}
We use metastable NaCl-structure Ti$_{0.5}$Al$_{0.5}$N alloys to probe effects of configurational disorder on adatom surface diffusion dynamics which control phase stability and nanostructural evolution during film growth. First-principles calculations were employed to obtain potential energy maps of Ti and Al adsorption on an ordered TiN(001) reference surface and a disordered Ti$_{0.5}$Al$_{0.5}$N(001) solid-solution surface. The energetics of adatom migration on these surfaces are determined and compared in order to isolate effects of configurational disorder. The results show that alloy surface disorder dramatically reduces Ti adatom mobilities. Al adatoms, in sharp contrast, experience only small disorder-induced differences in migration dynamics.
%The results show that alloy surface disorder gives rise to low-energy trapping sites, together with increased maximum energy barriers, for Ti adatoms, even along the most energetically favorable diffusion paths. Thus, the effect of disorder is to dramatically reduce Ti adatom mobilities. Al adatoms, in sharp contrast, experience only small disorder-induced differences in migration dynamics.

\end{abstract}

\maketitle

Thin film growth is a complex physical phenomenon controlled by the interplay of thermodynamics and kinetics. This complexity facilitates the synthesis of metastable phases, such as Ti$_{1-x}$Al$_{x}$N alloys, which are not possible to obtain under equilibrium conditions and broaden the range of available physical properties in materials design. Fundamental understanding of elementary growth processes, such as adatom diffusion, governing nanostructural and surface morphological evolution during thin film growth can only be developed by detailed studies of their dynamics at the atomic scale. Research has mostly been carried out using elemental metals, as reviewed in refs.~\cite{Jeong1999,Antczak2007}. Much less is known about the atomic-scale dynamics of compound surfaces, and particularly little about complex, configurationally disordered, pseudobinary alloys which are presently replacing elemental and compound phases in several commercial applications. % such as the low-index planes of TiN, which are highly anisotropic.% Kodambaka et al.~\cite{Kodambaka2006} have reviewed experimental and theoretical results for TiN(001) and TiN(111) surfaces. 

Kodambaka et al.~\cite{Kodambaka2000,*Kodambaka2002s,*Kodambaka2003,*Kodambaka2002b,*Kodambaka2006} and Wall et al.~\cite{Wall2004,Wall2005} used scanning tunneling microscopy to determine surface diffusion activation energies $E_s$ on both TiN(001) and TiN(111). However, due to the vast difference between experimental and adatom hopping time scales, determining diffusion pathways requires theoretical approaches via first-principles methods that are capable of providing clear atomistic representation on the ps time scale. Gall et al.~\cite{Gall2003surf} employed first-principles calculations to show that $E_{s}$ for Ti adatom diffusion on TiN is much lower on the (001) than the (111) surface and used this diffusional anisotropy to explain the evolution of (111) preferred orientation during growth of essentially strain-free polycrystalline films. The correspondingly large differences in chemical potentials result in Ti adatoms having higher residence times on (111) than on (001) grains. 
%However, much less is known about the surfaces of pseudobinary alloys which are presently replacing compounds in several industrial applications.
	
Here, we use cubic Ti$_{1-x}$Al$_{x}$N(001), a metastable NaCl-structure pseudobinary alloy, as a model system to probe the role of short-range disorder on cation diffusivities which control phase stability, surface morphology, and nanostructural evolution during growth. Ti$_{1-x}$Al$_{x}$N alloys with x $\sim 0.5$, synthesized by physical vapor deposition (PVD) far from thermodynamic equilibrium~\cite{Hakansson1987, *Adibi1991,*Greczynski2011}, are commercially important for high-temperature oxidation~\cite{McIntyre1990} and wear-resistant applications~\cite{Prengel1997,*PalDey2003,*Mayrhofer2003,Horling2005}. Alloying TiN with AlN has also been shown to alter surface reaction pathways controlling film texture and nanostructure~\cite{Horling2005,Beckers2005,Petrov1993,Adibi1993b}. Unfortunately, atomic-scale understanding of the growth of these important, and more complex, materials systems is presently rudimentary as best.
	Surface diffusion on a metal alloy, the CuSn system in ordered configurations and in the dilute limit~\cite{Chen2010PRL}, has only recently been considered using first-principles. However, it is well known that \emph{configurational disorder} can have large effects on the physical properties of solid solutions~\cite{Ruban2008REV}.

We employ first-principles calculations using the projector augmented wave method~\cite{Blochl1994} as implemented in the Vienna Ab-Initio Simulation Package (VASP)~\cite{Kresse1993}, to determine the energetics of cation adsorption and diffusion on ordered TiN(001) and conÞgurationally-disordered Ti$_{0.5}$Al$_{0.5}$N(001) surfaces. Electronic exchange correlation effects are modeled using the generalized gradient approximation~\cite{Perdew1996}. The plane wave energy cut-off is set to 400 eV. 
We sample the Brillouin zone with a grid of $3\times3\times1$ k-points.
%We use a Monkhorst-Pack scheme~\cite{Monkhorst1976} for sampling of the Brillouin zone with a grid of 3x3x1 k-points.

TiN(001), for reference, and Ti$_{0.5}$Al$_{0.5}$N(001) surfaces are modeled using slabs with four layers of $3\times3$ in-plane conventional cells with 36 atoms per layer. Calculated equilibrium lattice parameters, $a_0$, of bulk TiN, 4.255 \AA, and Ti$_{0.5}$Al$_{0.5}$N, 4.179 \AA, are employed. The vacuum layer above the surfaces corresponds to $5.5 a_0$. The adatoms are spin polarized, which is found to be important for Ti adatoms with its partially filled 3d-shell, but not for Al. 
	To investigate diffusion on a configurationally-disordered surface, the Ti$_{0.5}$Al$_{0.5}$N(001) slab is modeled using the special quasirandom structure (SQS) method~\cite{Zunger1990}. We impose a homogenous layer concentration profile and minimize the correlation functions on the first six nearest-neighbor shells for the slab as a whole.
%	 in which the metal-site correlation functions are minimized for the first six nearest-neighbor shells. %It should be noted that simulating the surface of a random alloy is a complex procedure for which, in principle, the composition distribution of the layers and the correlation functions, both within and between layers, should be considered individually~\cite{Marten2008}. However, in this work, we are primarily interested in qualitative effects of disorder and impose a homogenous layer concentration profile and minimize the correlation functions for the slab as a whole. 

%\begin{figure}[htb]
% 	\centering
%  	\includegraphics[width=0.95\columnwidth]{information}
%  	\caption{(Color online) (a) The atomic configuration of the top (001) surface layer of the SQS used to model adatom diffusion on Ti$_{0.5}$Al$_{0.5}$N(001). (b) Schematic description of the periodically repeated SQS adsorpition energy surfaces used as a first approach in the kinetic Monte Carlo simulation. (b) Schematics of the randomly rotated SQS adsorption energy surfaces used in the second approach.}
% 	\label{fig:information}
%\end{figure}

\begin{figure*}[htb]
 	\centering
  	\includegraphics[width=0.9\textwidth]{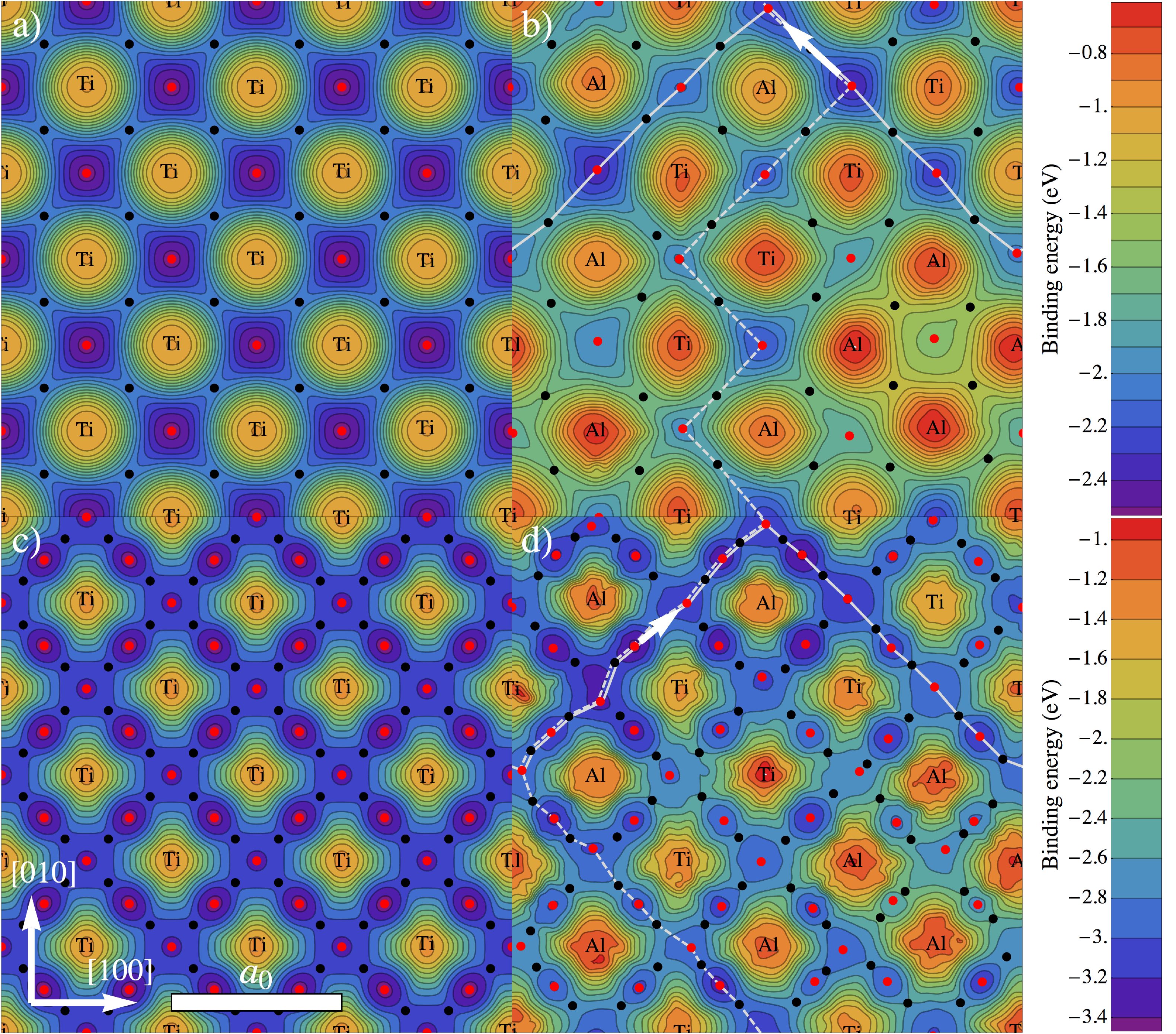}
  	\caption{(Color online) Adsorption energy surface for: (a) an Al adatom on TiN(001), (b) an Al adatom on Ti$_{0.5}$Al$_{0.5}$N(001), (c) a Ti adatom on TiN(001), and (d) a Ti adatom on Ti$_{0.5}$Al$_{0.5}$N(001). Local minima are marked with red dots, while black dots indicate saddle point barrier positions. White lines, solid and dashed, marks preferred paths for diffusion on the disordered surfaces.}
 	\label{fig:Esurf}
\end{figure*}

Convergence of diffusion barriers is tested with respect to the geometrical and numerical details of the calculations. $E_s$ results are within 0.04 eV of the converged value, partly due to error cancelation between the effects of treating Ti semicore states as core and the limited number of layers; both are of the order of 0.08 eV, but with opposite signs. Our primary focus is the observed differences in cation dynamics on the two surfaces.
%Convergence tests are performed for diffusion activation barriers with respect to the number of layers and in-plane size of the slab, the k-point grid, number of relaxed layers, vacuum thickness, and inclusion of semicore states in the valence. The barrier heights $E_s$ are within 0.04 eV of the converged value, partly due to error cancelation between the effects of treating Ti semicore states as core and the limited number of layers; both are of the order of 0.08 eV, but with different signs. Our primary focus is the observed differences in cation dynamics on the two surfaces.% Since the same computational approach is used to treat both surfaces, small systematic inaccuracies should, to a considerable extent, cancel out and not affect our conclusions.

We begin by calculating the adsorption energy $E^{Al,Ti}_{ads}(x,y)$ for Ti and Al adatoms as a function of positions x and y on both ordered TiN(001) and disordered Ti$_{0.5}$Al$_{0.5}$N(001) surfaces, 
\begin{equation}
E_{ads}^{Al,Ti}(x,y)=E_{slab+ad}^{Al,Ti}(x,y)-E_{slab}-E_{atom}^{Al,Ti}.
\end{equation}

\noindent $E^{Al,Ti}_{slab+ad}$is the energy of the slab with an adatom at $(x, y)$, $E_{slab}$ is the energy of the pure slab with no adatoms, and $E^{Al,Ti}_{atom}$ is the energy of an isolated Al or Ti atom in vacuum. We use a fine grid of sampling points, $\Delta x = \Delta y = 0.05a_0$. In each calculation, the adatom is fixed within the plane and relaxed out of plane. The upper two layers of the slab are fully relaxed, while the lower two layers are stationary. A periodic polynomial interpolation between the calculated points is used to obtain a smooth energy surface.

Adsorption-energy profiles for Al and Ti atoms on TiN(001) and Ti$_{0.5}$Al$_{0.5}$N(001) surfaces are shown in Figs. 1(a)-1(d). The most favorable sites for Al adatoms on both surfaces are directly above N atoms at bulk cation positions. For Al on TiN(001), Fig. 1(a), $E^{Al}_{ads}$ is -2.54 eV. On 
Ti$_{0.5}$Al$_{0.5}$N(001), Fig. 1(b), $E^{Al}_{ads}$ varies from -2.39 to -1.52 eV on the bulk cation sites depending on their local environment. Ti adatoms have two stable adsorption sites: fourfold hollows, surrounded by two N and two metal atoms, and the bulk site on-top N. For TiN(001), Fig. 1(c), $E^{Ti}_{ads}$ = -3.50 eV in the hollow site and -3.27 eV above N. On the alloy surface, Fig. 1(d), $E^{Ti}_{ads}$ varies from -3.42 to -2.58 eV in the hollow sites and -3.23 to -2.67 eV in on-top sites. Al-rich environments are much less favorable for both Al and Ti adatoms as can be seen in the lower right regions of Figs. 1(b) and 1(d). The overall preferred sites for Ti on Ti$_{0.5}$Al$_{0.5}$N(001) are fourfold hollow positions with one Ti and one Al nearest metal neighbors; not two Ti atoms as might have been expected. 

%%%%%%%%%%%%%%%%% K-MC
%%FLOW FIGURE
\begin{figure}[htb!]
 	\centering
  	\includegraphics[width=0.45\textwidth]{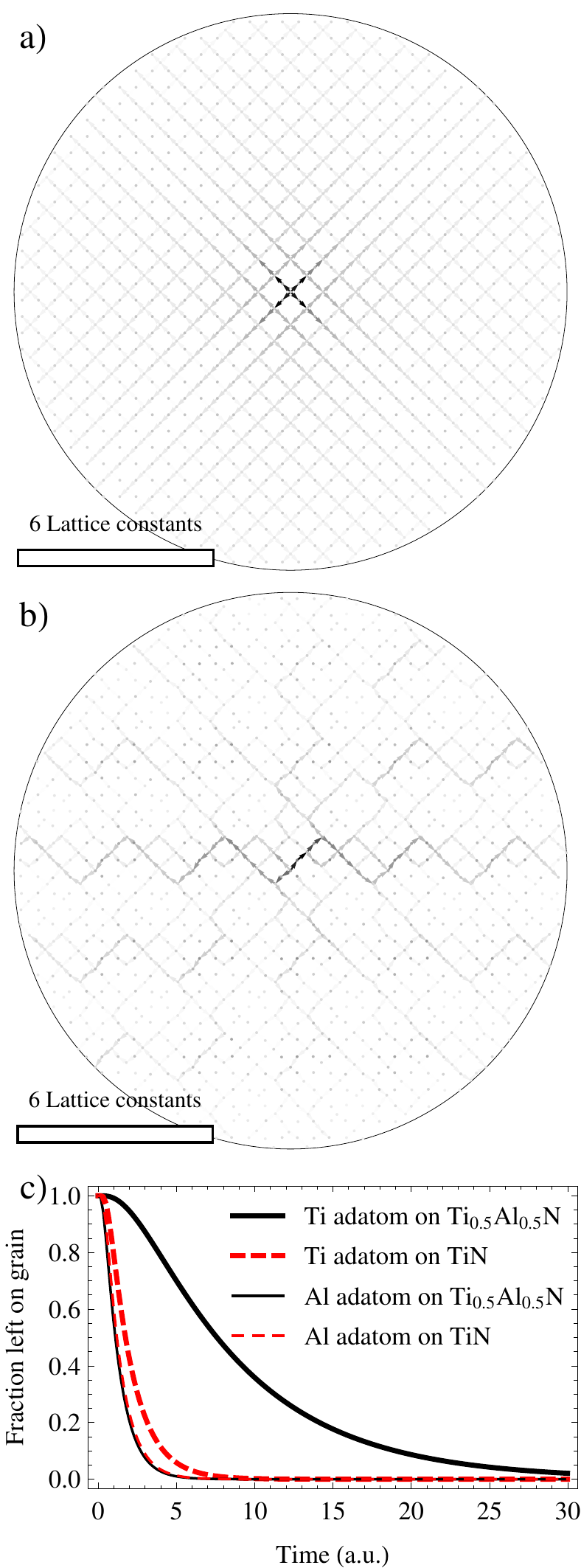}
  	\caption{ 
	(Color online) Ti adatom diffusion paths from the center to the edge of (001) surfaces of (a) TiN, and (b) disordered Ti$_{0.5}$Al$_{0.5}$N. (c) The probability as a function of time, that Al and Ti adatoms placed at random positions in the center of a circular grain of radius $8.5 a_{0}$ have not yet reached the grain boundary on TiN(001) and Ti$_{0.5}$Al$_{0.5}$N(001).}
 	\label{fig:flowfig}
\end{figure}

In order to quantify the impact of disorder on diffusion, transition state theory within a kinetic Monte Carlo approach is used to determine the mobilities of independent adatoms. The probability at each time step for a Ti or Al adatom at site $i$ to jump to site $j$ is calculated as

%In order to quantify the impact of disorder on diffusion, we use a kinetic Monte Carlo approach and the transition state theory to simulate the mobility of independent adatoms. On all the energy surfaces the probability in each timestep for an adatom at local minima site $i$ to jump to site $j$ is calculated as

\begin{equation}\label{eq:prob}
\Gamma_{ij}=\nu_0~ \mathrm{exp} \left( \frac{-\Delta E_{ij}}{k_B T} \right)
\end{equation}

\noindent where $\Delta E_{ij} = \left( E_{ij} - E_i \right)$ is the difference between the adsorption energy in the local minima $i$ and at the saddle point defining the barrier height $E_{ij}$ between sites $i$ and $j$. The temperature $T$ is 800~K, a representative value for PVD growth of transition-metal nitride thin films. For convenience, we choose the attempt frequency $\nu_0$ to be the same for Ti and Al on both TiN(001) and Ti$_{0.5}$Al$_{0.5}$N(001) surfaces, but note that Al adatoms should have a slightly higher attempt frequency than Ti due to their lower mass. %Since we are only interested here in individual adatom diffusion, and not adatom interactions, 
We determine $n_i(t)$, the probability density of finding adatoms on a given site $i$, at time $t$,  corresponding to an ensemble average of a large number of individual cases.

The most probable Al and Ti diffusion paths are identified by imposing a constant probability density of adatoms at the centers of circular grains with radii $8.5 a_0$, and then propagating the probability density using Eq.~\ref{eq:prob}. Adatoms crossing a grain boundary are not allowed to cross back. Thus, we obtain an adatom probability flow between sites i and j from the center of the grain outward,

\begin{equation}
F_{ij}= n_i\Gamma_{ij}-n_j\Gamma_{ji}.
\end{equation}

Steady-state results are plotted for Ti adatoms in Fig. 2 as grayscale intensity proportional to $F_{ij}$. Panel 2(a) shows that the flow of Ti adatoms across the ordered TiN(001) surface is symmetric and utilizes all $[110]$ paths. However, the flow of Ti atoms across Ti$_{0.5}$Al$_{0.5}$N(001), panel 2(b), simulated using periodically repeated SQSs, is almost completely absent in the energetically least favorable regions; most diffusion takes place along special paths. Such paths are indicated in Fig. 1(d) (Fig. 1(b) for Al adatoms) by white solid and dashed lines corresponding approximately to connections among the most favorable local energy minima. %, crossing the supercell surface horizontally and vertically.

Next, we determine the timescales of adatom diffusion on the two  surfaces. Fig. 2(c) is a plot of the probability as a function of time that adatoms, individually placed at a randomly chosen site close to the center of a circular grain, have not yet reached the grain boundary. On the pure TiN(001) surface, Al and Ti adatoms show similar behavior as the somewhat higher barriers for Al diffusion are compensated by Ti adatoms having three times as many local minima positions for a constant grain size. The striking result, however, is that Ti adatoms diffuse much slower on Ti$_{0.5}$Al$_{0.5}$N(001) than on TiN(001), while the rates for Al adatoms on the two surfaces are nearly equal. Since both Ti and Al adatoms diffuse predominantly along preferential paths on the disordered TiAlN(001) surface, the mobility differences are, in large part,  explained by differences in energy profiles along these paths. 

%%%%%%%%%%%%%% Analysis 2D barriers

\begin{figure}[htb!]
 	\centering
  	\includegraphics[width=0.7\textwidth]{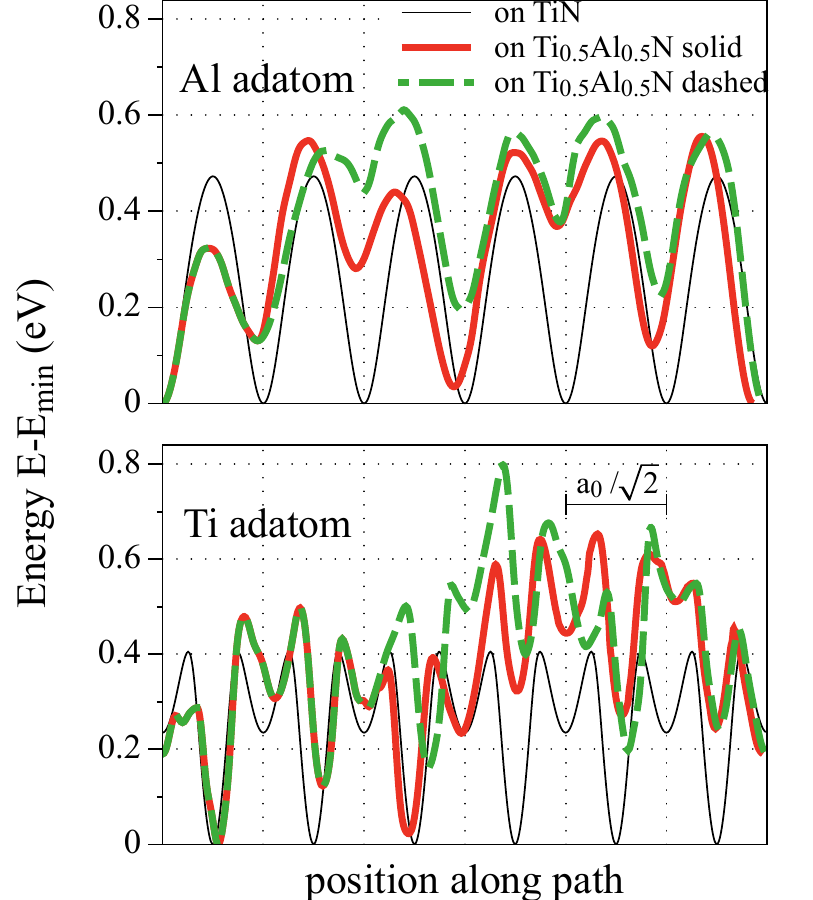}
  	\caption{(Color online) Adsorption energies of Al (upper graph) and Ti (lower graph) adatoms along favorable diffusion paths on ordered TiN(001) and disordered Ti$_{0.5}$Al$_{0.5}$N(001) surfaces. For the disordered alloy surface, energy profiles are plotted for both the solid and dashed paths across the SQS  shown in Fig.~\ref{fig:Esurf}.}
 	\label{fig:bar2D}
\end{figure}

Fig. 3 contains plots of $E^{Al,Ti}_{ads}$ relative to the most favorable adsorption site, along the preferred  diffusion paths on Ti$_{0.5}$Al$_{0.5}$N indicated in Figs. 1(b) and 1(d). The arrows in Fig. 1 define the starting position for the energy-path plots in Fig. 3. Corresponding $E^{Al,Ti}_{ads}$ plots on TiN(001) are included for comparison. The calculated Al adatom diffusion activation energy on TiN(001) is $E_s = 0.47$ eV. Both the solid and dashed low-energy paths for Al on Ti$_{0.5}$Al$_{0.5}$N(001) exhibit the signature of configurational disorder with alternating deep and less-deep energy minima. However, the individual barrier heights are, in most cases, considerably lower on the disordered surface with the maximum barrier height just $1.2\times$ larger than on TiN(001). $E_s$ for Ti adatoms on TiN(001) is 0.40 eV and the smaller barrier for jumping out of the minima atop N is 0.17 eV. The individual barriers for Ti on Ti$_{0.5}$Al$_{0.5}$N(001) are similar, but a series of less favorable energy minima, combined with asymmetric jump probabilities, creates additional migration obstacles approximately 2/3 along the outlined paths. The maximum obstacles are $2.0\times$ and $1.6\times$ higher than $E_s$ on TiN(001) for the dashed and solid diffusion paths, respectively, explaining the dramatic reduction of mobility in this case. The mass difference between Al and Ti atoms (which we ignored in these calculations) affects $\nu_0$ and will further increase the mobility difference.

These results illustrate the complex effects that configurational disorder can induce on surface diffusion. They also help to understand aspects of the growth behavior of Ti$_{1-x}$Al$_{x}$N thin films. Our observed increase in the residence time of Ti adatoms on Ti$_{0.5}$Al$_{0.5}$N(001) vs. TiN(001) is consistent with the experimentally reported transition in texture for polycrystalline TiN films, grown at relatively low temperatures with little or no ion irradiation, from (111)~\cite{Greene1995} toward (001) upon alloying with AlN~\cite{Horling2005}. In addition, the higher mobility of Al, with respect to Ti, adatoms on Ti$_{1-x}$Al$_{x}$N explains the results of Beckers et al. showing  AlN enrichment in (111) and Al depletion in (001) oriented grains~\cite{Beckers2005}.

In conclusion, we have compared the adsorption-energy landscape and the migration mobilities of Ti and Al adatoms on ordered TiN(001) and disordered Ti$_{1-x}$Al$_{x}$N(001) surfaces. The configurational disorder on the alloy surface results in the formation of deep trap sites for Ti adatoms which, together with an asymmetric adsorption energy map, dramatically decreases the Ti adatom mobility. In contrast, Al adatom mobilities are nearly the same on TiN(001) and disordered Ti$_{1-x}$Al$_{x}$N(001) surfaces due to a much smaller disorder-induced spread in energy minima values and more symmetric diffusion probability distributions along the most favorable paths on the alloy surface. These results explain observed differences in preferred orientation and nanostuctural evolution during growth of polycrystalline TiN and Ti$_{1-x}$Al$_{x}$N films.

\begin{acknowledgments}
We acknowledge financial support by the Swedish Foundation for Strategic Research (SSF), the Swedish Research Council (VR), and the European Research Council (ERC). The simulations were carried out using supercomputer resources provided by the Swedish national infrastructure for computing (SNIC).

%	We gratefully acknowledge financial support by the Swedish Foundation for Strategic Research (SSF) Program in Materials Science for Nanoscale Surface Engineering, MS$^2$E, the Swedish Research Council (VR), and teh European Research Council (ERC). The simulations were carried out using supercomputer resources provided by the Swedish national infrastructure for computing (SNIC).
\end{acknowledgments}

%textwidth in cm: \printinunitsof{cm}\prntlen{\textwidth}\\
%textwidth in cm: \printinunitsof{cm}\prntlen{\linewidth}

%%\bibliography{/Users/bjoal/Documents/Fysik/CrAlN/coatings}
%\bibliography{/Users/bjoal/Documents/Fysik/CrAlN/coatings.bib}

%\bibliography{bibl}


\end{document}